\documentclass[]{interact}

\usepackage{epstopdf}
\usepackage{endnotes}
\usepackage[numbers,sort&compress,merge]{natbib}
\renewcommand\bibfont{\fontsize{10}{12}\selectfont}

\newcommand{\avg}[1]{\langle #1 \rangle}

\newcommand{\ket}[1]{| #1 \rangle}

\newcommand{\ketbra}[2]{| #1 \rangle\!\langle #2 |}

\begin{document}


\title{Why material slow light does not improve cavity-enhanced atom detection}

\author{
\name{B. Megyeri$^\ast$, A. Lampis$^\ast$\thanks{$^\ast$These authors contributed equally to this work.}, G. Harvie, R. Culver, and J. Goldwin\thanks{CONTACT J. Goldwin. Email: j.m.goldwin@bham.ac.uk}}
\affil{School of Physics \& Astronomy, University of Birmingham, Edgbaston, Birmingham B15 2TT, UK}
}

\maketitle

\begin{abstract}
We discuss the prospects for enhancing absorption and scattering of light from a weakly coupled atom in a high-finesse optical cavity by adding a medium with large, positive group index of refraction. The slow-light effect is known to narrow the cavity transmission spectrum and increase the photon lifetime, but the quality factor of the cavity may not be increased in a metrologically useful sense. Specifically, detection of the weakly coupled atom through either cavity ringdown measurements or the Purcell effect fails to improve with the addition of material slow light. A single-atom model of the dispersive medium helps elucidate why this is the case.
\end{abstract}


\section{Introduction}

The sensitivity of weak optical scattering measurements can be improved by coupling the atoms or molecules under test to a high-finesse optical cavity \cite{Ye03}. The confined geometry increases the intensity per photon, and the recirculation of light increases the effective length of the medium. For many cavity-based techniques, the detection sensitivity is proportional to the resonator quality factor, $Q$, which can be as high as $10^{11}$ for an optical cavity with kilohertz linewidth. Typically one tries to increase the quality factor either by reducing the losses or increasing the cavity length. The former strategy is currently limited by the technical challenge of reducing scattering losses below the ppm level, while the latter compromises the inherent advantage of confinement. It is therefore of interest to explore new methods for augmenting traditional cavity-enhanced measurements.

One possibility is to add a dispersive element to the cavity, in order to influence the light scattering dynamics. In particular, the technique of electromagnetically induced transparency (EIT) can be used to generate a nearly transparent medium with extremely large group index and correspondingly small group velocity \cite{EIT}. The presence of slow light in a cavity can narrow the cavity transmission spectrum \cite{Lukin98, Wang00}, and reduce the decay rate \cite{Laupretre11}. It seems natural then to ask whether the resulting increase in the $Q$-factor is metrologically useful. Despite the apparently straightforward nature of this question, the answer is not as obvious as it may seem.

Recently a consensus has emerged that not all slow light media behave the same, leading to a distinction between \emph{structural} and \emph{material} slow light \cite{Mork10,Boyd}. Optical resonators themselves are an example of structural slow light media, with all of the advantages mentioned above. Photonic crystals and fibre Bragg gratings are other examples. Within structural slow light media, constructive interference increases the amplitude of the electric field, leading to an enhancement of light-matter interactions. In contrast, EIT represents material slow light. For such media, the total energy density increases, but the light intensity remains the same. In an elegant experiment, a direct comparison showed that structural slow light increased Beer-Lambert absorption, but material slow light did not \cite{Thevenaz12}.

Here we theoretically investigate combining structural and material slow light in the application of two specific types of cavity-enhanced metrology. In cavity ringdown spectroscopy, a cavity is filled with a variable density of gas under test; the cavity is probed with a laser and weak absorption in the gas increases the decay rate of the light after abrupt extinction of the probe \cite{Ye03}. In the second method, the gas is excited directly by a probe laser propagating in a direction transverse to the cavity axis; light is scattered into the cavity via the Purcell effect \cite{Purcell}, and detected upon transmission. In the absence of dispersion, the sensitivity of both types of measurement can be characterised by the cooperativity, $C=g^2/(\kappa\gamma)\propto Q$, where $g$ is the matrix element coupling the light and atoms, and $2\kappa$ and $2\gamma$ are the energy decay rates of the cavity and atomic excited state, respectively. As we discuss in more detail below, the cavity decay rate is inversely proportional to the group index \cite{Soljacic05,Sauvan05}, implying a potential improvement with slow light. However the coupling $g$ decreases as the square root of the group index for material slow light, as can be seen by considering the energies in the dielectric medium and the electromagnetic field \cite{Boyd,Bradshaw11}. This suggests the cooperativity is independent of the group index. We show that $C$ remains the key parameter for these techniques in the presence of material slow light, and therefore no improvement occurs.

The rest of this paper is organised as follows. In Section \ref{sec:cavity} we model the slow light medium as a single atom undergoing EIT inside a high finesse optical cavity. The effective group index is calculated, and the effects on the cavity linewidth and lifetime are discussed. In Section \ref{sec:CRD} we add a two level atom and consider a cavity ringdown measurement. We show that slow light fails to increase the influence of the atom on the cavity decay rate. In Section \ref{sec:Purcell} we show that Purcell scattering similarly is not improved with the addition of slow light. The scattering rate from the atom is roughly unaffected, and the potential benefit of the increased cavity lifetime is negated by a suppression of the photonic excitation. Finally we conclude with some remarks on how to realise our model experimentally.

\section{\label{sec:cavity}Cavity linewidth and lifetime}

Quantum optical theories of dispersive dielectrics have been investigated at least since the 1940s, with considerable progress made in the early 1990s \cite{Watson49,Drummond,Glauber91,Huttner92,Philbin10}. The development of a rigorous theory of such media runs into interesting subtleties involving causality and losses, which are outside the scope of the present work. Instead we offer a single-atom model based on electromagnetically induced transparency. The advantage of this model is that simple analytic expressions can be obtained. The practical realisation of the model is discussed in the conclusion.

We first consider the case with only the slow-light medium in the cavity. The slow light effect arises from a single strongly-coupled thee-level atom (3LA), undergoing electromagnetically induced transparency. The energy levels are in a ladder or $\Xi$ configuration, with states $\ket{G}$, $\ket{E}$, and $\ket{D}$, in ascending order. The upper state $\ket{D}$ is assumed to have a relatively long lifetime, as with high-lying Rydberg states, to allow the formation of a narrow transparency feature with large dispersion. The cavity is tuned so that the empty cavity resonance frequency matches the free-space electric dipole transition $\ket{G}\leftrightarrow\ket{E}$, and the $\ket{E}\leftrightarrow\ket{D}$ transition is resonantly addressed by a coupling laser for EIT. Finally, the cavity is probed by a weak laser assumed to be in a coherent state. In the rotating wave approximation, and in a frame rotating at the probe laser frequency, the Hamiltonian is ($\hbar=1$),
\begin{eqnarray}\nonumber
H &=& -\delta(a^\dagger a + \sigma_{EE} + \sigma_{DD}) -i\eta(a - a^\dagger) \\ &~&\quad -G(\sigma_{GE}\,a^\dagger + \sigma_{GE}^\dagger\,a) -\tfrac{1}{2}W(\sigma_{DE}+\sigma_{DE}^\dagger) \quad.
\label{eq:H3LA}
\end{eqnarray}
The operator $a$ annihilates a cavity photon, and the $\sigma_{ij}=\ketbra{i}{j}$ operators act on the atomic states; $\delta$ is the probe detuning, $\eta$ is the probe amplitude (proportional to the square root of incident power), $G$ is half the Rabi frequency between the 3LA and a single photon, and $W$ is the coupling beam Rabi frequency, which controls the effective group index of the 3LA through the EIT effect. We have neglected kinetic energy, as appropriate for a sufficiently cold trapped atom.

\begin{figure}
\centering
\includegraphics[height=6cm]{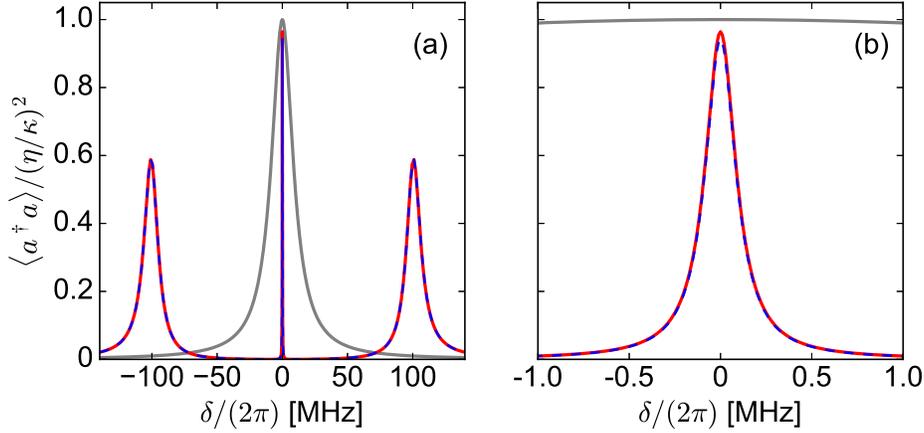}
\caption{Cavity spectrum with EIT. The normalised cavity photon number is plotted as a function of probe detuning. The solid red curve is the solution of the master equation, the dashed blue curve is the approximation from Eqs.(\ref{eq:rho3LA}) and (\ref{eq:alpha3LA}), and the grey curve is for the empty cavity. The parameters were $(G,W,\kappa,\gamma)=2\pi\times(100,20,10,3)$~MHz, $\eta=\kappa/1000$, and $\Gamma=\gamma/1000$, as defined in the text. (a): full spectrum across all three normal modes. (b): central feature due to EIT.}\label{fig1}
\end{figure}

To describe decay due to cavity transmission and losses, and spontaneous emission from $\ket{E}$ and $\ket{D}$, we use the master equation for the density operator \cite{EIT}. We assume that expectation values of products of atomic and field operators are separable, and the cavity field is in a coherent state with complex amplitude $\alpha=\avg{a}$. The equation of motion for the cavity field is,
\begin{eqnarray}
\dot{\alpha} &=& \eta -(\kappa - i\delta)\,\alpha + iG\rho_{EG} \quad.
\label{eq:alphadot3LA}
\end{eqnarray}
Here $\kappa$ is the field decay rate, and $\rho_{EG}=\avg{\sigma_{GE}}$. As pointed out for two-level atoms in \cite{Zhu90}, we can use the result for $\rho_{EG}$ which one would obtain in free space \cite{EIT}, but with a probe Rabi frequency of $2G\alpha$. To first order in $\alpha$,
\begin{eqnarray}
\rho_{EG} &=& G\alpha\,\frac{\delta+i\Gamma}{(W/2)^2-\delta^2+\gamma\Gamma-i\delta(\gamma+\Gamma)} \quad,
\label{eq:rho3LA}
\end{eqnarray}
with $\gamma$ and $\Gamma$ equal to half the spontaneous emission rate from $\ket{E}\to\ket{G}$ and from $\ket{D}\to\ket{E}$, respectively. For simplicity we assume $\ket{D}\to\ket{G}$ is forbidden.\endnote{What matters is that the lifetime of $\ket{D}$, including all decay channels, is long. Including $\ket{D}\to\ket{G}$ decay does not qualitatively change our results.} Substituting this expression into Eq.(\ref{eq:alphadot3LA}) and solving for steady state ($\dot{\alpha}=0$),
\begin{eqnarray}
\alpha &=& \frac{\eta}{\kappa - i\delta - iG\rho_{EG}/\alpha} \quad.
\label{eq:alpha3LA}
\end{eqnarray}
Note that $(\rho_{EG}/\alpha)$ is independent of $\alpha$, as a consequence of the assumption of weak probing in the linear regime, implicit in Eq.(\ref{eq:rho3LA}). The spectrum of in-cavity photon number $\avg{a^\dagger a}=|\alpha|^2$ can now be obtained. An example is shown in Fig.~\ref{fig1}(a), for $(G,W,\kappa,\gamma)=2\pi\times(100,20,10,3)$~MHz, and $\eta=\kappa/1000$ and $\Gamma=\gamma/1000$. These parameters satisfy $\sqrt{\gamma\Gamma}\ll(W/2)\ll G$, as required for negligible absorption and large group index in the EIT medium \cite{EIT,Lukin98}. The spectrum exhibits a narrow central transparency window and two relatively broad side peaks. The latter correspond to the usual normal modes of a two-level atom in the cavity \cite{Zhu90}, shifted to slightly larger detunings by the added dispersion. The approximation based on Eq.(\ref{eq:alpha3LA}), which is shown as the dashed curve, is nearly indistinguishable from the full numerical solution of the master equation (solid curve) obtained with the QuTiP software package \cite{QuTiP1,QuTiP2}.\endnote{The computer code used to produce all of the figures in this work is available online at {\ttfamily https://arxiv.org/abs/1705.01028}.}

The central transparency feature, highlighted in Fig.~\ref{fig1}(b), is the main focus of this work. First we consider a cavity ringdown (CRD) measurement, where the probe laser is instantaneously turned off ($\eta\to 0$), allowing the cavity field to decay freely. To describe the decay at long times, we Taylor-expand $\rho_{EG}$ to first order around $\delta=0$,
\begin{eqnarray}
\dot{\alpha} &\approx& \eta - (\kappa' - i\delta')\,\alpha \quad,
\end{eqnarray}
which takes the same form as an empty cavity, with effective detuning $\delta'=n_g\,\delta$ and decay rate $\kappa'=\kappa+\kappa_{3\rm LA}$, where we have introduced,
\begin{eqnarray} \label{eq:ng}
n_g &=& 1+\left(2G/W\right)^2 \\ \label{eq:kappa3LA} 
\kappa_{3\rm LA} &=& (n_g - 1)\,\Gamma \quad.
\end{eqnarray}
Because $\rho_{EG}$ is proportional to the linear susceptibility \cite{EIT}, Eq.(\ref{eq:ng}) identifies $n_g=101$ as the effective group index of the 3LA, and Eq.(\ref{eq:kappa3LA}) shows the effect of residual absorption due to imperfect transparency, with $\kappa_{3\rm LA}\ll\kappa$. Now the steady-state solution is just $\alpha=\eta/(\kappa'-i\delta')$; viewed as a linear response function, there is a complex pole $\delta_*=0-iR/2$ given by the solution of $\kappa'-in_g\delta_*=0$. The parameter $R$ gives the energy decay rate at long times,
\begin{eqnarray}
R &=& 2\,\frac{\kappa'}{n_g} \quad.
\label{eq:R3LA}
\end{eqnarray}
This shows the competition between residual absorption in the 3LA, which increases the decay rate, and the group index, which decreases it \cite{Lukin98}. This was studied experimentally with inhomogeneously broadened atoms in \cite{Wang00}. For our parameters, the effect of absorption is small and the group index is large, so that $R\approx(2\kappa/n_g)\ll 2\kappa$. Thus the cavity linewidth is narrowed and the photon lifetime increased \cite{Laupretre11}, each by a factor of approximately $n_g$. 

\begin{figure}
\centering
\includegraphics[height=6cm]{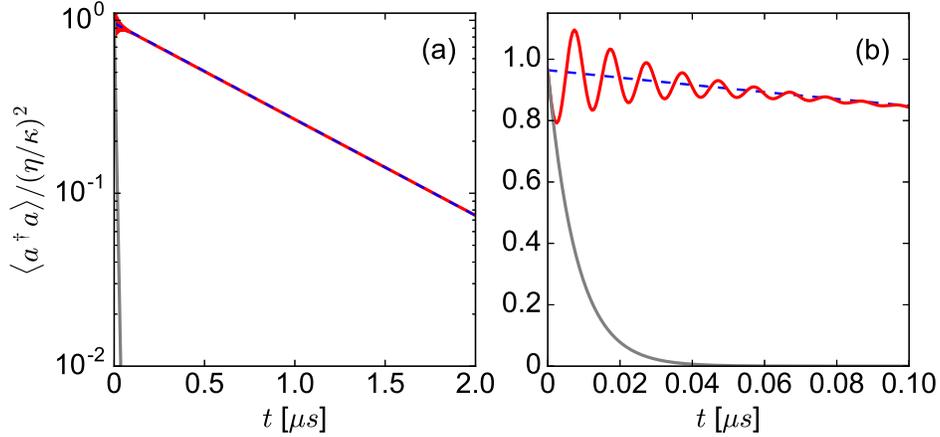}
\caption{Cavity ringdown with EIT. All parameters are the same as in Fig.~\ref{fig1}, with $\delta=0$ and $\eta\to 0$ at time $t=0$. The red curve is the full solution of the master equation, the blue dashed curve is an exponential decay with rate given by Eq.(\ref{eq:R3LA}), and the grey curve is for an empty cavity. (a): Full ringdown trace (note the logarithmic scale). (b): Transient oscillation for early times.}\label{fig2}
\end{figure}

Figure~\ref{fig2} shows the simulated cavity ringdown corresponding to the spectrum in Fig.~\ref{fig1}. The increased lifetime is immediately apparent when compared to the case without the 3LA in Fig.~\ref{fig2}(a). In Figure~\ref{fig2}(b) we see an initial transient oscillation with the 3LA, which is due to the excitation of the side modes when the cavity probe is suddenly extinguished. This is reminiscent of the Rabi oscillations observed with two-level atoms when the probe power is abruptly changed \cite{Brecha95}. Here the oscillation frequency is $\sim G$ instead of $2G$, representing beating between the central mode and the two side modes. These transients decay at a rate of order $(\gamma+\kappa)\gg R$.

Given the narrowing of the cavity spectrum and the increase in the lifetime, it is conventionally said that the cavity quality factor $Q$ increases by a factor of $n_g$ \cite{Soljacic05,Sauvan05}. However although the lifetime increases, the number of round trips does not. In the next sections we will see that the extended interaction time alone fails to improve detection of a weakly coupled two-level test atom in two exemplary cases.

\section{\label{sec:CRD}Cavity ringdown detection}

We now add to the system a single two-level atom (2LA), which is weakly coupled to the cavity field. The Hamiltonian becomes,
\begin{eqnarray}
H &\mathrel{{+}{=}}& -\delta\sigma_{ee}-g(\sigma_{ge}\,a^\dagger + \sigma_{ge}^\dagger\,a) \quad,
\label{eq:H3LA2LA}
\end{eqnarray}
where $2g$ is the Rabi frequency for the 2LA and a single photon. The ground and excited states of the atom, $\ket{g}$ and $\ket{e}$, can be the same as the corresponding states of the 3LA, as long as the 2LA is not exposed to the EIT coupling beam. If the transition is the same, we can assume the 2LA is located at a low-intensity point of the cavity mode, where $g\ll G$, so that it does not strongly modify the spectrum described above. The equation of motion for $\alpha$ becomes,
\begin{eqnarray}
\dot{\alpha} &=& \eta -(\kappa - i\delta)\,\alpha + iG\rho_{EG} + ig\rho_{eg} \quad,
\label{eq:alphadot3LA2LA}
\end{eqnarray}
where $\rho_{eg}=\avg{\sigma_{ge}}$ is obtained from the free-space expression as,
\begin{eqnarray}
\rho_{eg} &=& -g\alpha\,\frac{\delta-i\gamma}{\delta^2+\gamma^2} \quad.
\label{eq:rho2LA}
\end{eqnarray}
We have restricted ourselves again to first order in $\alpha$ in Eq.(\ref{eq:rho2LA}). Continuing as before, we obtain the ringdown rate,
\begin{eqnarray}
R &=& 2\,\frac{\kappa' + g^2/\gamma}{n_g - g^2/\gamma^2} \quad.
\label{eq:R3LA2LA}
\end{eqnarray}
The term proportional to $g^2$ in the numerator reflects the added absorption due to the 2LA. In analogy with the above we could define $\kappa_{2\rm LA}=C\kappa$, where $C=g^2/(\kappa\gamma)$ is the cooperativity. The negative term proportional to $g^2$ in the denominator describes the small, anomalous dispersion of the 2LA \cite{Shimizu02,Lien16}. This term is neglected in conventional CRD measurements \cite{Ye03}, where $\kappa\ll\gamma$, but plays an observable role for the parameters we consider here. Finally, we define a figure of merit for CRD detection of the 2LA by considering the relative effect of $C$ on the decay rate,
\begin{eqnarray}
\mbox{FOM} &\equiv& \frac{R(C)-R(0)}{R(0)} \quad.
\label{eq:FOM}
\end{eqnarray}
For $C\ll 1$, we expect $\mbox{FOM}\propto C$. 

\begin{figure}
\centering
\includegraphics[height=6cm]{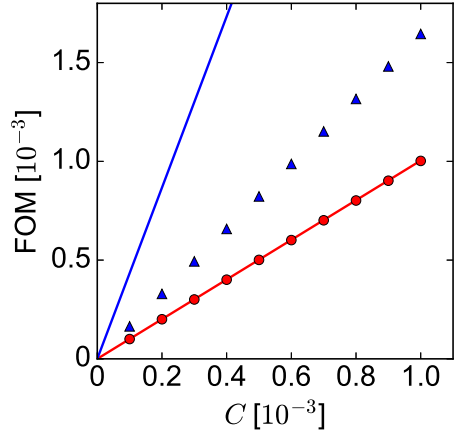}
\caption{Figure of merit for cavity ringdown, as a function of 2LA cooperativity. Red circles/blue triangles show FOM from Eq.(\ref{eq:FOM}) obtained from numerical simulations with/without the 3LA slow light, where simulated CRD signals were fit to exponential decays at long times, ignoring initial transients. The 3LA slow light effect used the parameters of Figs.~\ref{fig1} and \ref{fig2}. Solid lines are the predictions from Eq.(\ref{eq:R3LA2LA}).}\label{fig3}
\end{figure}

Figure~\ref{fig3} shows $\mbox{FOM}$ for the 2LA with and without the slow light effect. It is clear that the 3LA actually reduces the figure of merit (red circles) compared to the case with only the 2LA (blue triangles). The slow light results obey $\mbox{FOM}\approx C$, as one would have for a conventional CRD measurement with $n_g=1$ and $\kappa\ll\gamma$. Without the 3LA, the negative group index of the 2LA plays a signifcant role, although one which is overestimated by Eq.(\ref{eq:R3LA2LA}). We find that this prediction always improves when $\gamma\gg R/2$, whether $n_g$ is large or not, which we attribute to the limitations of the first-order Taylor expansion of $\rho_{ge}$ around $\delta=0$. When the cavity linewidth exceeds $\gamma$, higher-order terms in the dispersion temper the superluminal effect. Viewed in the time domain, if the excited state lifetime is not short compared to the cavity lifetime, then $\avg{\sigma_{ee}}$ acts as a slowly decaying source of light for the cavity which creates a bottleneck in the ringdown. In any event, the numerical simulations make it clear that slow light extends the time scale of the CRD measurement without improving its sensitivity. This holds true for all of the combinations of parameters we have tested, with large group indices and small residual absorption.

To summarise, cavity ringdown measurements fail to improve with the addition of material slow light.\endnote{One potential exception is if there is an intrinsic advantage to slowing down the ringdown time scale, regardless of any effects on the absorption. An example was considered theoretically in \cite{Yang04}.} Such measurements benefit from the number of round trips of light in the resonator, which increases the effective absorption length of the medium. Although the cavity $Q$-factor increases with large $n_g$, the number of round trips --- inversely proportional to the losses --- does not. As mentioned in the introduction, the cooperativity $C$ is independent of the group index for material slow light, and $C$ remains the essential parameter characterising cavity ringdown under these conditions. We show in the next section that this still only tells a part of the story.

\section{\label{sec:Purcell}Purcell effect}

We now turn to the Purcell effect, whereby direct laser excitation of the two-level atom leads to scattering into the cavity mode. We thus take $\eta=0$ and add to the Hamiltonian,\begin{eqnarray}
H &\mathrel{{+}{=}}& -\tfrac{1}{2}w(\sigma_{ge} + \sigma_{ge}^\dagger) \quad,
\label{eq:HPurcell}
\end{eqnarray}
where $w$ is the Rabi frequency of the driving laser. In the absence of dispersion, the atom scatters into the cavity mode with a rate equal to $C\gamma=g^2/\kappa$ \cite{Purcell,Kleppner}. This can be considered an application of Fermi's golden rule, where the factor of $g^2$ represents the square of the transition matrix element, and $1/\kappa$ is the density of photonic states. Because this light populates a single cavity mode, Purcell scattering can offer an efficient way to detect extremely weak transitions, with an enhancement proportional to the cavity $Q$-factor. If the cooperativity is unchanged in the presence of material slow light, then we might expect the same for the scattering rate. If so, the detection rate will actually decrease, as the cavity emission rate becomes $2\kappa'/n_g$.

\begin{figure}
\centering
\includegraphics[height=6cm]{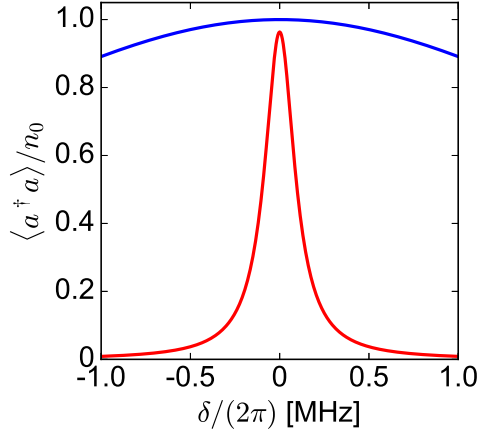}
\caption{Purcell effect with (red, narrow) and without (blue, broad) slow light. The parameters were the same as in Figs.~\ref{fig1} and \ref{fig2}, with $g=2\pi\times 0.1$~MHz and $w=\gamma/1000$. The scaling factor, $n_0=(wg)^2/(2\kappa\gamma)^2$, is the same for both curves.}\label{fig4}
\end{figure}

The master equation for $\rho_{eg}$ reads,
\begin{eqnarray}
\dot{\rho}_{eg} &=& -(\gamma-i\delta)\rho_{eg} + i(g\alpha + \tfrac{1}{2}w)(\rho_{gg}-\rho_{ee}) \quad.
\label{eq:mastereqrhoeg}
\end{eqnarray}
If the two-level atom is far from saturation, $\rho_{gg}\approx 1$ and $\rho_{ee}\approx 0$. For zero detuning, and assuming $\kappa^{-1}\dot{\alpha}\ll 1$ and $(W/2)^2\gg \gamma\Gamma$,
\begin{eqnarray}
\dot{\rho}_{eg} &\approx& i\,\frac{w}{2} - \left(\gamma + \frac{g^2}{\kappa'}\right)\rho_{eg} \quad.
\label{eq:Purcell}
\end{eqnarray}
This shows that the scattering rate, now $g^2/\kappa'$, is approximately unchanged, in agreement with the notions that $Q\to n_g Q$ and $g\to g/\sqrt{n_g}$. Given the increased lifetime, the in-cavity photon number should increase. Figure~\ref{fig4}, comparing in-cavity spectra for Purcell scattering with and without the 3LA, shows that this is not the case. Both spectra are normalised to the same number $n_0=(wg)^2/(2\kappa\gamma)^2$. The slow light narrows the Purcell spectrum without increasing the peak number of photons. If the scattering rate has not changed, and the cavity lifetime has increased, then why does the in-cavity number of photons remain the same?

To understand how this happens, we neglect all decay channels and focus on the Hamiltonian evolution. With $\delta=0$ and $w=0$, we treat the interaction term in Eq.(\ref{eq:H3LA2LA}) as a perturbation to the Hamiltonian $H_0$ in Eq.(\ref{eq:H3LA}). We restrict ourselves to the subspace of states with only a single excitation, $\{\ket{0,G,e}, \ket{1,G,g}, \ket{0,E,g}, \ket{0,D,g}\}$, where numbers denote Fock states of the field. The eigenstates of $H_0$ in this subspace include the two side modes near $\pm G$ in Fig.~\ref{fig1}(a), and two degenerate modes around zero --- one is predominantly $\ket{0,D,g}$ with a small admixture of $\ket{1,G,g}$, and represents the central transparency feature, and the other is just $\ket{0,G,e}$, which contributes no energy when $g=0$. Applying degenerate perturbation theory, we find the eigenstates and energies to first order in $g/G$. Then with $\ket{0,G,e}$ as the initial condition, we calculate how the excitation of the 2LA is transferred to the strongly coupled 3LA-cavity complex,
\begin{eqnarray}
\label{eq:Pee}
\avg{\sigma_{ee}} &\approx& \cos^2\left(\tfrac{gt}{\sqrt{n_g}}\right) \\
\label{eq:Pn}
\avg{a^\dagger a} &\approx& \tfrac{1}{n_g}\,\sin^2\left(\tfrac{gt}{\sqrt{n_g}}\right) \\
\label{eq:PEE}
\avg{\sigma_{EE}} &\approx& 0 \\
\label{eq:PDD}
\avg{\sigma_{DD}} &\approx& \left(1-\tfrac{1}{n_g}\right)\sin^2\left(\tfrac{gt}{\sqrt{n_g}}\right) \quad.
\end{eqnarray}

We can now clarify some key aspects of the dynamics. First, the excited state of the 2LA is most directly coupled to the upper EIT state; only a small fraction $(1/n_g)$ of the initial excitation energy is taken up by the cavity field. Furthermore the Rabi frequency shows the expected $g\to g/\sqrt{n_g}$ behaviour. These two effects are related to the storage of energy in a macroscopic dielectric \cite{Boyd}, which can be expressed as a re-scaling of the electric field $\mathcal{E}\to\mathcal{E}/\sqrt{n_g}$ \cite{Bradshaw11}. Finally, the extended cavity lifetime can be attributed to the predominantly $\ket{0,D,g}$ nature of the central spectral feature. The lifetime of this state, $1/(2\Gamma)$, is much longer than the bare cavity lifetime, allowing the 3LA to act as a long-lived phase memory, as described previously in the context of superradiant lasing \cite{Bohnet12}.

Combining everything, we find that material slow light affects Purcell scattering as follows. The coupling of the 2LA to the cavity field is strongly reduced. The in-cavity number of photons is approximately restored by the prolonged cavity lifetime. But the flux of photons out of the cavity is reduced again. As before, we conclude that material slow light fails to improve the measurement.

\section{Conclusion}

We have considered the effects of material slow light when combined with measurements of cavity ringdown and Purcell scattering. A simple model was introduced, with dispersion provided by a single three-level atom subject to electromagnetically induced transparency. Although the slow light was shown to increase the quality factor of the cavity, the light-matter cooperativity was unchanged. As such the absorption and scattering from the atom under test were not enhanced. With respect to the Purcell effect, slow light presented a significant detriment. This was traced back to the transfer of energy to the weakly radiating state of the three-level atom, at the expense of reduced excitation of the cavity field. 

We reiterate that these conclusions only apply to material slow light. We expect that coupling atoms to structural slow light in cavities and waveguides will find numerous applications along the lines of what we have studied here (see, for example, \cite{Goban14,Zang16} and references therein). Furthermore there are other scenarios where dispersion engineering, either material or structural, can be advantageous. Structural slow light can be  used to enhance both absorption and gain \cite{Mortensen07,Ek14}, and to improve conventional phase interferometers \cite{Shi07}, Fourier transform interferometers \cite{Shi07FT}, and grating spectrometers \cite{Shi13}. Anomalous dispersion may improve cavity-based inertial sensors \cite{Wicht97,Salit10,Smith16}, and active sensors (i.e., operating above the lasing threshold) can be enhanced through a variety of fast- and slow-light schemes \cite{Shahriar07,Bohnet12,Weiner12,Kotlicki12,Scheuer15}.

The \emph{single-atom dielectric}, which we have considered here for simplicity, can be realised in current state-of-the-art experiments with cold atoms in high finesse microcavities. Experiments on single-atom detection of rubidium atoms have reached sufficiently small cavity mode volumes and low losses to achieve the cavity parameters we have assumed \cite{Trupke07,Gehr10,Biedermann10,Goldwin11}. Rydberg state lifetimes of $1000/(2\gamma)$ have been measured for $n{\rm D}_{5/2}$ states of rubidium atoms in a magneto-optical trap \cite{Branden09}, providing realistic upper levels for the $\Xi$-EIT scheme. Alternatively, $\Lambda$-type level schemes \cite{Mucke10,Kampschulte10} could be employed with qualitatively similar results. A cavity length of $\sim 100~\mu$m is sufficient to allow individual addressing of the two atoms and to suppress Rydberg blockade effects \cite{Jaksch00,Lukin01}. Future work will involve extending the single-atom model to include anomalous dispersion for zero and negative group indices \cite{Bradshaw11,Laupretre12}.

\section*{Acknowledgements}

This work evolved from discussions at the workshop, ``Quantum sensors for precision positioning and underground mapping," November, 2013, at the Instituto de F\'{i}sica de S\~{a}o Carlos, Universidade de S\~{a}o Paulo, funded by FAPESP and the Universities of Birmingham and Nottingham. We are grateful to Philippe Courteille, John Weiner, Celso Jorge Villas-Boas, Vincent Boyer, and Peter Krueger for useful discussions, and to Dan Gauthier and Selim Shahriar for pointing out numerous metrologies which benefit from engineered dispersion. B.~Megyeri is supported by DSTL (DSTLX1000092132), and all others by EPSRC. We thank Giovanni Barontini for feedback on the manuscript.

\theendnotes

\end{document}